\documentclass[prl,aps,showpacs,twocolumn,floatfix,amsmath,amssymb]{revtex4}
\usepackage{graphicx}
\usepackage{ulem}
 \def\ba#1{\begin{array}{#1}}
 \def\ea{\end{array}}
 \def\be{\begin{equation}}
 \def\ee{\end{equation}}
\def\bq{\begin{equation}}
\def\eq{\end{equation}}
 \def\br{\begin{eqnarray}}
 \def\er{\end{eqnarray}}

 \def\lsim{\:\raisebox{-0.5ex}{$\stackrel{\textstyle<}{\sim}$}\:}

\begin{document}

\title{ Core polarization, Brown-Rho scaling\\
and a memory of Gerry's Princeton Years }
\author{ T.\ T.\ S.\ Kuo   }
\affiliation{Department of Physics, State University New York, 
Stony Brook, New York 11794, USA}	
	
\author{ J.\ W.\ Holt   }
\affiliation{Department of Physics,  University of Washington,
Seattle, WA 98195, USA}	
	
\date{\today}
\begin{abstract}
Core-polarization (CP) and Brown-Rho (BR) scaling were 
among Gerry's most favorite topics.
In this contribution, we will discuss some of the early history 
as well as more recent work associated 
with these two fascinating phenomena. I (TTSK) will begin with
some recollections of Princeton, where I met Gerry
for the first time in 1964 and worked as his postdoc. Core polarization 
was in fact the first topic he assigned to me.  JWH started
working with Gerry at Stony Brook from 2003 and 
was Gerry's last student in nuclear physics.
We three had teamed up well, working closely
on both CP and BR scaling, particularly on the latter's connection to 
the anomalously-long beta-decay lifetime of carbon-14. 
We shall here briefly review these topics, including a recently
developed new Brown-Rho scaling based on a Skyrmion half-Skyrmion
two-phase model.
\end{abstract} 
\maketitle

\section{ I. Introduction}
This is a tribute to Gerry Brown who was our mentor, collaborator 
and good friend for a long long time, 
from 1964 to 2013 for TTSK, and from 2003 to 2013 for JWH. We remember him 
and our times together very well, and here we shall review briefly 
two subjects, `core 
polarization' and `Brown-Rho scaling', on which we have 
collaborated extensively over the years. 
Before doing so, we think we should first describe a 
recent book project \cite{brownkuobook} we worked on 
together:

Gerry talked with me (TTSK) one day in early 2007 about writing 
a book on `Nucleon-Nucleon Interactions
and the Nuclear Many-Body Problem'. Gerry's idea was to put together 
a reprint volume with a few 
introductory chapters and a collection of our published
works spanning a period of about forty years. I of course was very 
happy about the idea, and soon afterward World-Scientific (WS) 
agreed to support the project. Gerry 
then asked JWH (then finishing up his Ph.D with Gerry) 
and Prof.\ Sabine Lee (Univ.\ of Birmingham) to help with 
the writing, typsetting (LaTexing) and organization of the book.
(Gerry often said ``The young are supposed to help the old.'')
As usual, Gerry was a man of quick action. In the summer of 2007 we
 had meetings at Gerry's home, 
and a tentative plan of `who-does-what' was laid. 
So we started to work. Gerry in fact quickly wrote
many neatly hand-written pages and sent us copies of them. 
Gerry often said he had only a 5-dollar 
pocket calculator, a sort of excuse he used to 
avoid doing calculations on computers and learning to use LaTex. 
So Sabine nicely typset all of Gerry's hand-written notes, and 
attached below is what Gerry wrote, in his 
unique style, in a memorable preface for the book:

\vskip 0.2cm

{ {\bf Preface: Why now is a good time to write about the
Nucleon-Nucleon Interaction and the Nuclear Many Body Problem}}
\vskip 0.1cm
``Why do two old nuclear physicists, with the help of a
junior colleague and a historian, now write about the nucleon-nucleon
interaction to which they have devoted such a large portion of their
research lives previously?

The immediate explanation is straightforward. The main problems at
the level of meson exchange physics have been solved. We now have an
effective nucleon-nucleon interaction $V_{low
- k}$, pioneered in a renormalization group formalism by several
of us at Stony Brook and our colleagues at Naples, which is
nearly universally accepted as the unique low-momentum interaction
that includes all experimental information to date.

Why does this make reconstructing the history of our
understanding of the nucleon-nucleon interaction necessary or
useful? There are several good reasons for engaging in a historical
appreciation of the progression of research and the developments
leading to our current knowledge in this subject area.

First, our understanding is based on a multi-step development in which a
variety of different scientific insights and a wide range of
physical and mathematical methodologies fed into each other. This is
best appreciated by a looking at the different `steps along the
way', starting with the pioneering work by Brueckner and
collaborators, which was just as
necessary and important as the insightful, masterly improvements to
Brueckner's approach by Hans Bethe and his students.
The main achievement in the work of Brueckner and Bethe
et al.\ was the `taming' of the hard core of the nucleon-nucleon potential, 
which has since been understood to result from the exchange of the
$\omega$-meson, a `heavy' photon.
The off-shell effects which bedevilled Bethe's work
that ended up in the 1963 Reference Spectrum Method
 were treated
relatively accurately by introducing an energy gap between initial
bound states and intermediate state. Kuo and Brown showed that this
would be accurately handled by taking the intermediate states to be
free; i.e.\ by just using Fourier
components, as now done in the effective field theory resulting from
the renormalization group formalism.

Well, one can say to the young people that this is `much ado about
nothing'. In fact, long ago, when Gerald E.\ Brown was Professor at 
Princeton, Murph Goldberger (turning on its head Winston Churchill's famous
quote about the R.A.F.\ during the Battle of Britain) 
claimed in reference to the nuclear interaction that `never have so
many contributed so little to so few.' Admittedly, at the time it
was hard going.

If we had a unique set of interactions, one for each angular
momentum, spin and isospin channel, it could be argued that it would be
justified to stop there. However, since Brueckner came on the scene,
Bethe reorganized the theory, Kuo and Brown wrote their paper that
prepared the effective field theory by using the Scott-Mozskowski
separation method, and chiral invariance hit the scene. Chiral
invariance does not do anything for Yukawa's pion exchange, because
the pion gets most of its mass from somewhere outside of the low-energy
system, maybe by coupling to the Higgs boson. But the masses of the
other mesons drop with increasing density, like
$$
m_\rho^* \cong m_\rho (1-0.2 n/n_0) \hspace{2cm} \mbox{``Brown/Rho
scaling''}
$$
where $n$ is the density and $n_0$ is nuclear matter saturation density.
The change in masses of the scalar-$\sigma$ and vector-$\omega$ mesons
pretty much cancel each other in effects--the scalar exchange giving
attraction and the vector repulsion. However, in the tensor force,
the $\rho$-exchange `beats' against the pion exchange, the former
cancelling more and more of the latter as the density increases.
This decrease with density of the tensor force interaction has
important effects:
\begin{enumerate}
\item It is responsible for saturation in the nuclear many-body
system.
\item It converts an around hour-long carbon-14 lifetime from a
superallowed transition in the Wigner $SU(4)$ for $p$-shell nuclei
into an archeologically long 5,700-year transition.
\end{enumerate}
`Brown/Rho scaling' is also important for neutron stars
 and may play an important role in turning
them into black holes and for `cosmological natural
selection'.
It must be admitted that the same effects could be given by
three-body forces, but Brown/Rho scaling has a deep connection
with chiral symmetry restoration. We shall review these facets in detail.

Undoubtedly, much more is to come, but we believe that now is a good
time to summarize the interesting history of the nucleon-nucleon
interaction.''
\vskip 0.2cm

Indeed Gerry has devoted a large portion of his research life
to nuclear physics, especially to questions related 
to the nucleon-nucleon interaction
and nuclear many-body problem. He has also devoted 
a large portion of his life to guiding, 
helping and taking care of his students, postdocs and colleagues
(including both of us). 
Gerry had two operations in 2008, and was not in 
good health afterwards. With Gerry ill, 
we worked hard together with Sabine to finish the book, which was 
published by WS in early 
2010. Many of Gerry's friends and colleagues visited him 
regularly while he was recuperating. 
I (TTSK) and my wife Annette also visited him often (about once or more
each month), and it happened that we saw him in the afternoon 
of May 27, 2013, just four days before he passed away. 
He was particularly cheerful that 
afternoon, smiling, tasting a pastry and making a 
typical Gerry-style joke. In the following, let
us describe briefly the two research projects we have
worked on together.

\section{II. Core polarization}

In September 1964, I (TTSK) went to Princeton as an instructor (which
is a research associate with minor teaching duties) to work with
Gerry. I only learned much later that Gerry knew my advisors
Elizabeth Baranger (Univ.\ of Pittsburgh) and Michel Baranger
(Carniege-Mellon Univ.) very well, and that they had arranged for me
to work with Gerry. In fact, they were all close associates of Hans Bethe.
I went to the Palmer Physical Laboratory one day and met
Gerry for the first time. I still remember well when he introduced
Chun-Wa Wong to me and said ``let me introduce my secret 
weapon to you'' (so I realized from that first interaction 
that Gerry had a good
sense of humor and liked to make jokes). Chun-Wa was also a research 
associate of Gerry; he was a graduate student at Harvard and completed
his thesis with Gerry in Copenhagen.

I (JWH) had a similar experience.
I met Gerry for the first time in the spring of 2003 when I was a first-year 
graduate student at Stony Brook University. 
My brother Jason (who was already working 
in the nuclear theory group with TTSK) introduced me one day to Gerry, 
whose first words to me came as a surprise: ``Is your middle name William?''
After I answered ``yes'', Gerry smiled and said ``My grandson's name is Jeremy 
William. Why don't you come work with me this summer?'' That was the easiest 
job interview of my life, but the challenging part came 
later as I tried to keep
up with Gerry's diverse research interests, from hadronic physics, 
to nuclear astrophysics, to low-energy nuclear structure theory.
Soon afterwards, I began to focus on BR scaling \cite{holt04}
and CP \cite{holtkbb07}.

At Princeton Gerry had a rather large and active nuclear
theory group. Senior faculty members were Gerry and Ben Bayman.
To the best of my (TTSK) memories, the research associates were 
(in alphabetical order, here and later)
   J.\ Blomquist,
   J.\ Flores,
   W.\ Friedman,
   A.M.\ Green,
   A.\ Kallio,
   T.T.S.\ Kuo,
   H.\ Picker,
   A.\ Lande,
   P.\ Mello,
   G.\ Ripka,
   C.W.\ Wong, and
   L.\ Zamick.
Visiting faculty members were
   A.\ Arima,
   L.\ Castillejo,
   I.\ Talmi,
   H.\ McManus,
   M.\ Moshinsky,
   H.\ Lipkin and
   P.\ Zilser.
Gerry had about ten graduate students during his four-year stay
at Princeton. I only remember a few of them, namely
   G.\ Bertsch,
   M.Y.\ Chen,
   W.\ Gerace, 
   H.\ Mavromatis, 
   J.\ Noble and
   I.\ Sharon.
I remember Gerry once said ``Bertsch was too fast: I gave him a problem
and he would disappear for a couple of weeks and come back with
the solution. So I soon ran out of problems. I let him graduate.''
Princeton also had a very active nuclear experimental group, which
consisted of (as far as I remember) R.\ Sherr, J.\ McCullen, O.\ Ames
and G.\ Garvey.

Gerry's nuclear theory group worked closely with the Rutgers
nuclear physics group, which was a large group with G.\ Temmer, A.\ Covello,
G.\ Sartoris and others. Every Monday
afternoon we went to Rutgers (about a 20-mile drive from Princeton)
to attend their weekly seminar. Every Thursday evening they came
to Princeton's `bull session', a Gerry specialty.
Usually we all went to have dinner together, and then came to
the seminar room at about 7 pm, starting the bull session
which was an informal seminar with lots of discussions
and `no-time-limit'. Typically it ran for about 3 hours or more
till about 11 pm. In those years, computers were still
`primitive'; we used card punchers to punch cards
and submit jobs (boxes of computing cards) at the computing
center. (Gerry often mentioned that in his graduate-student days
the computers used paper tapes as inputs.)
So after the bull session, the `young' postdocs
almost all first drove to the computing center and submitted some jobs
before going home.

The so-called Kuo-Brown matrix elements 
\cite{kuobrown66,brownkuo67,kuobrown68} were first developed at that
time in order to provide a microscopi basis for the nuclear shell
model. We shall only briefly describe them, as more detailed discussions
about them have been given by Osnes and by Coraggio (see contributions
by them in this memorial volume). The NN interaction and the nuclear 
many-body problem are both difficult problems. Gerry recently
wrote in his book \cite{brownkuobook}:

``One of the authors, Gerry Brown, arrived at Princeton in early 
September, 1964. The next morning,
as he came to the Palmer Physics Laboratory, Eugene Wigner, who just preceded
him, opened the door for him (It was a real contest to get ahead of
Eugene and open the door for him which very few succeeded in doing.)
Eugene asked Gerry, as he went into the building, what he planned to
work on. `I plan to work out the nucleon-nucleon interaction in
nuclei.' Eugene said that it would take someone cleverer than him,
to which Gerry replied that they probably disagreed what it meant to
`work out'. Gerry wanted to achieve a working knowledge, sufficiently
good to be able to work out problems in nuclear physics...''

It indeed turned out to be very hard to `work out'
the nucleon-nucleon (NN) interaction in nuclei
in a fundamental way, and a more feasible and 
physically-motivated approach is to compute instead
an `effective' or `renormalized' nucleon-nucleon interaction. 
After I (TTSK) arrived Princeton, Gerry asked me
to study Brueckner theory \cite{brownkuobook}, which was a new and difficult
subject for me at that time. I am still indebted to 
Chun-Wa for helping me greatly in learning
the theory, which was originally designed for nuclear matter 
but which Gerry intended to apply to finite 
nuclei in a shell model approach.

Consider as an example the nucleus $^{18}O$ which 
has 18 nucleons. In the shell-model effective theory
for this nucleus, it is reduced from a many-body problem 
with `eighteen' nucleons to
a simplified one of just `two' valence nucleons residing 
outside of an inert $^{16}O$ core.
This is a simple, smart and bold step, and I think it is of 
the type of physics that Gerry appreciated. But the 
two nucleons are in fact renormalized quasi-nucleons, which are 
different from the original bare ones.
The interaction for bare nucleons is $V_{NN}$ while that for the
quasi-nucleons is $V_{eff}$. We first tried $V_{eff}=G$ 
where $G$ is the Brueckner G-matrix and used it to
calculate the low-lying spectra of $^{18}O$ and $^{18}F$.
The calculated spectra were however in poor agreement with 
experiment, so Gerry then suggested that we 
take the effective interaction as
\begin{equation}
V_{eff}=G+G_{3p1h},
\end{equation}
where $G$ represents the direct interaction of the valence nucleons
via a $G$-matrix interaction. The term $G_{3p1h}$ denotes
the second-order core-polarization diagram shown in Fig.\ 1(a).

It was to our great joy that the inclusion of 
$G_{3p1h}$ greatly improved our results. 
For example, it significantly lowered the lowest $0^+$
as well as raised a group of high-lying states of $^{18}O$, making
the calculated spectrum in good agreement with experiment.
  The matrix elements based on $G+G_{3p1h}$ are generally referred to
as the Kuo-Brown (KB) matrix elements 
\cite{brownkuobook,kuobrown66,brownkuo67,kuobrown68} and 
have been widely used in
nuclear shell model calculations for decades with remarkably
successful results (see for example Refs.\ \cite{poves81,wild84,brown88}).

NN interactions are short-ranged as is the $G$-matrix interaction. 
In the 1960s it
 was found mainly in Copenhagen (led by Bohr and Mottelson) that 
to describe empirical nuclear properties we need 
also a long-range effective $P_2$-force
of the form \cite{brownkuobook,ringschuck80}
\begin{equation} 
V_{P_2}=-\chi \Sigma _{ij} {r_i^2} {r_j^2} 
Y_{2,m}(\theta _i,\phi_i)
Y_{2,-m}(\theta _j,\phi_j)(-1)^m.
\end{equation} 
This empirical force was in fact well reproduced by the core-polarization
diagram $G_{3p1h}$ \cite{brownkuo67}, which allowed 
two nucleons far from each other
to interact indirectly through excitations of the core. 
I (TTSK) remember well a cartoon-like picture
of the core polarization effect drawn by Gerry and Akito Arima, 
where two satellites orbit near the 
earth surface. They are far away from each other 
on opposite sides of the earth, so that they can 
hardly interact with each other directly. But they can interact 
with each other
via the tidal waves induced by them.

\begin{figure}
\includegraphics[width=7cm]{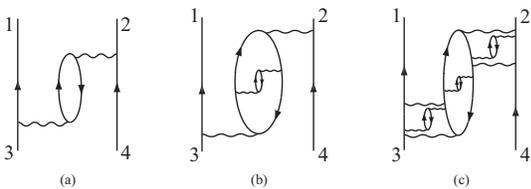}
\caption{Core polarization diagrams.}
\label{fig.1}
\end{figure}

The success of the Kuo-Brown interactions has led to a 
number of further studies \cite{barrettk,jensen95,corag09}, but 
since the KB core polarization diagram is only a second-order one, 
a natural question remained: how significant are the 
higher-order diagrams?

This is a very important question, and 
we (the Holt brothers, former student Scott Bogner, TTSK and Gerry)
have indeed made extensive efforts in answering it as reported in 
Ref.\ \cite{holtkbb07}. In comparison with our 
earlier calculation \cite{brownkuobook}, we have made several
improvements: (i) We employ the renormalization group (RG) 
low-momentum nucleon-nucleon interaction 
$V_{low-k}$ \cite{bogner01,kuo02,bogner02,kuo0102,schwenk02,bogner03,holt04},
(ii) folded diagrams \cite{kuoosnes,klr} are summed 
to all orders, and (iii) an induced-interaction approach is used where
particle-particle and particle-hole vertex functions are 
calculated self-consistently \cite{holtkbb07}.

Microscopic nuclear many-body calculations using 
realistic $V_{NN}$ interactions 
are complicated by the difficulties caused by strong 
repulsive cores normally found in such interactions.
For many years, a standard procedure to overcome such difficulties
has been the Brueckner $G$-matrix method, where $V_{NN}$ is converted to
a smooth $G$-matrix effective interaction by summing 
ladder diagrams to all orders in 
the nuclear medium. However, in many ways $G$ is not 
convenient for many-body calculations.
First, its Pauli exclusion operator is complicated for 
calculation, and second $G$ is energy dependent 
in an off-energy-shell manner.
%For example, to evaluate the matrix element 
%$\langle k_1 k_2|G(\omega)|k_3k_4\rangle$ for a certain vertex in a diagram 
%we need to know the energy variable $\omega$.  But knowing  the external
%indices $(k_1,k_2,k_3,k_4)$ alone is in general not adequate to
% determine $\omega$;
%we need to have  also information about  what other particles 
%in the diagram are doing.
These features  complicate the calculation of diagrams
 with the $G$-matrix interaction, 
especially for high-order diagrams such as diagrams (b) and (c) of Fig.\ 1.

The low-momentum NN interaction $V_{low-k}$ is based on a renormalization 
group approach where one integrates out momentum components beyond a decimation 
scale $\Lambda$
\cite{bogner01,kuo02,bogner02,kuo0102,schwenk02,bogner03,holt04}.
Briefly speaking, it is given by a pair of $T$-matrix equivalence relations:
\begin{equation}
  T(k',k,k^2) = V_{NN}(k',k)   
 + P\int _0 ^{\infty} q^2 dq \frac{V_{NN}(k',q) T(q,k,k^2)} 
 {k^2-q^2 }, 
\end{equation} 
\begin{eqnarray}
 && T(p',p,p^2) =   V_{low-k}(p',p) \nonumber \\ 
  &&+ P\int _0 ^{\Lambda} q^2 dq  \frac{V_{low-k}(p',q)T (q,p,p^2)}
 {p^2-q^2 }, 
  ~(p',p) \leq \Lambda 
\end{eqnarray}
where $P$ denotes the principal value integral. From the above equations, $V_{low-k}$ can be obtained from $V_{NN}$.
Note that $V_{low-k}$ is energy independent and thus convenient for many-body calculations. 
There are a number of high precision models for $V_{NN}$ \cite{paris,bonnabc,bonns,cdbonn,argonne,nijmegen,idaho}, but their
$\langle k |V_{NN}|k'\rangle$ matrix elements are in fact significantly different from each other although they all reproduce
the experimental two-nucleon data quite well \cite{bogner03}. An amazing feature of the different $V_{low-k}$  derived
from the above different $V_{NN}$ potentials is that they are nearly identical to each other for $\Lambda \lesssim 2.0 fm^{-1}$, leading to
a nearly universal low-momentum NN interaction \cite{bogner03}. Realistic NN potentials are all constructed to fit the experimental 
NN phase shifts up to $E_{lab} \leq ~350$ MeV which corresponds to 
$\Lambda \simeq 2 fm^{-1}$, providing an explanation for why
$V_{low-k}$ with this decimation scale should be nearly universal.

In our new CP calculation \cite{holtkbb07}, we employed a folded-diagram expansion 
\cite{kuoosnes,klr} which provides a formally exact method for calculating the effective interaction $V_{eff}$
for valence nucleons outside a closed core. It is of the form
\begin{equation}
V_{eff} = \hat{Q} - \hat{Q'} \int \hat{Q} + \hat{Q'} \int \hat{Q} \int 
\hat{Q} - \hat{Q'} \int \hat{Q} \int \hat{Q} \int \hat{Q} + ~...~~,
\end{equation}
where each $\int$ symbol represents a `fold'. Each $\hat Q$-box
represents a collection of irreducible diagrams as shown
by the diagrams of Fig.\ 1. The $\hat Q'$-box is the same as $\hat Q$-box
except it starts from the second-order diagrams, namely
$\hat Q'=\hat Q - V_{NN}$.

As is well known, high-order CP calculations are difficult to perform, largely because the number of
diagrams grows rapidly as one goes to higher orders in perturbation theory. The number of diagrams at third 
order is already quite large, though still manageable \cite{barrettk,jensen95,corag09,corag12}, but it was soon 
realized that an order-by-order calculation of CP diagrams beyond third order is not practicable. 

To fully assess the effects of core polarization to high order, a non-perturbative method is called for. 
The non-perturbative method we use is based on the elegant and rigorous induced interaction 
approach of Kirson \cite{kirson} and Babu and Brown \cite{babu}, hereafter referred to as
KBB.  Other successful non-perturbative summation methods have been
developed, such as the parquet summation \cite{jackson} and
the coupled cluster expansion \cite{dean}. 
In the KBB  formalism the vertex functions are obtained
by solving a set of self-consistent equations, thereby generating
CP diagrams to all orders such as diagrams (b) and (c) of Fig.\ 1. 

\begin{figure}
\includegraphics[width=2.8in,angle=270]{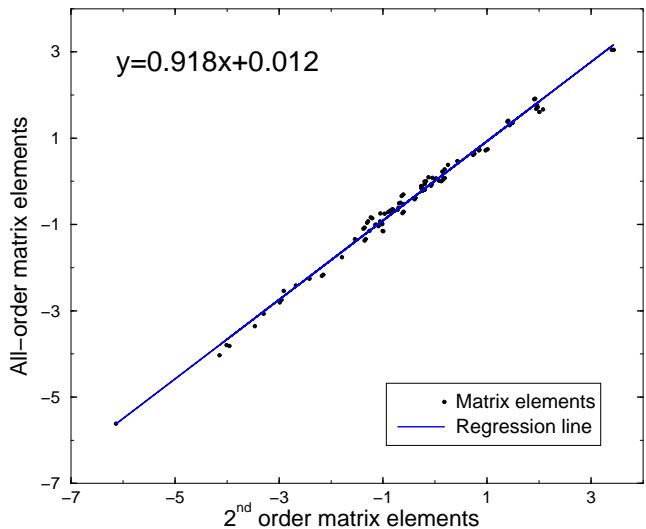}
\caption{A comparison of the second-order core 
polarization matrix elements with those of the all-order KBB calculation.} 
\label{fig.3}
\end{figure}
\begin{figure}
\includegraphics[height=2.4in,width=2.8in]{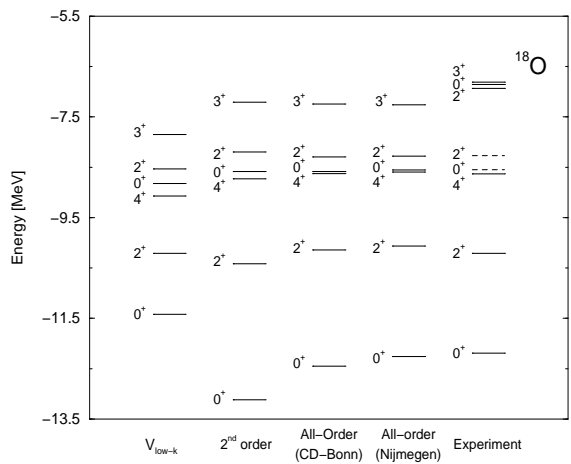}
\caption{Spectra for the $^{18} \rm O$ system calculated to different orders
  in perturbation theory. Dashed lines for the experimental levels
\cite{tilley} indicate levels with large intruder
  state mixing \cite{wild84,brown88}. }
\label{fig.4}
\end{figure}
  Let us now give a brief summary of our all-order calculation
for the shell-model effective interaction $V_{eff}$. We first use $V_{low-k}$ to 
calculate the $\hat Q$-box. And in this $\hat Q$-box  the bubble-in-bubble
CP diagrams, like those shown in Fig.\ 1, are included to all orders in a
self-consistent way.
Then we obtain $V_{eff}$ by summing up the $\hat Q$-box folded-diagram
series of Eq.\ (5).
To illustrate our results, we show in Fig.\ 2 a comparison of 
the $sd$-shell second-order CP matrix elements with those
given by the all-order KBB calculation.  The two groups
of matrix elements are rather close to each other with the all-order
elements being about $10 \%$ weaker. A similar comparison for the $^{18}O$ spectra is given in Fig.\ 3. 
The spectrum given in the `$V_{low-k}$' column is obtained with
the $\hat Q$-box composed of the first-order diagram only, and consequently
the resulting spectrum is too compressed compared to experiment. 
The spectrum given by the `2$^{\rm nd}$ order' column is obtained with
the $\hat Q$-box composed of the first- and second-order diagrams.
The  inclusion of the KBB CP diagrams in the $\hat Q$-box largely
improves the agreement of the resulting spectra (labeled `all-order') with
experiment.

\section{III. Brown-Rho scaling}

Gerry moved to Stony Brook in 1968 and set up a large and 
very active nuclear theory group. Faculty members in his group
were initially Akito Arima, Andy Jackson and TTSK. There were
indeed a large number of visitors and postdocs during Gerry's first
years at Stony Brook. Gerry took very good care of them, often
putting them up in his home (so that, as Gerry would say, `they
will work day and night'). As far as TTSK can remember, the visitors
and postdocs included 
    S.\ Backman, 
    D.\ Bes,
    J.\ Blomqvist, 
    R.\ Broglia,
    M.\ Chemtob,
    K.\ Dietrich, 
    J.\ Durso, 
    P.\ Ellis,
    A.\ Fessler, 
    B.\ Friman,
    H.\ Gayer,
    E.\ Hajimichael,
    G.\ Hering,
    M.\ Ichimura, 
    L.\ Ingber,
    M.\ Kawai,
    D.\ Kurath, 
    R.\ Lawson,
    H.K.\ Lee,
    G.L.\ Li,
    Z.X.\ Li,
    Z.Y.\ Ma,
    R.\ Machleidt,
    H.\ Muether,   
    E.\ Nyman,
    F.\ Osterfeld, 
    E.\ Oset,
    E.\ Osnes,
    H.\ Pauli,
    Dan-Olaf Riska,
    M.\ Rho, 
    J.P.\ Shen,
    R.\ Silbar,
    H.Q.\ Song,
    J.\ Speth, 
    D.\ Strottman,
    K.\ Suzuki, 
    J.\ Vergadoes,
    N.\ VinhMau, 
    R.\ VinhMau, 
    J.\ Wambach,
    W.\ Weise, 
    H.F.\ Wu,
    S.S.\ Wu,
    S.D,\ Yang,
    Z.Y.\ Zhang ...  

Having a large number of physicists working together
was very pleasant and productive, leading to many long-term
collaborations. Mannque Rho was
a frequent visitor, and it was at Stony Brook where 
Brown-Rho scaling \cite{brownrho91,brownrho04} originated.  

Realistic nuclear potentials are mediated by the  exchange
of mesons such as the $\pi$-, $\rho$-, $\omega$- and $\sigma$-meson.
In constructing these potentials the meson-nucleon coupling constants
are adjusted to fit the `free-space' NN scattering data. 
Mesons in a nuclear medium, however,
can have properties (masses and couplings) that are 
different than in free space, as the former
are `dressed' or `renormalized' by their interactions with the medium.
Thus, the NN potential in medium, denoted by $V_{NN}(med)$, should be
different from that in free-space.

How to obtain $V_{NN}(med)$ is of course a most difficult and challenging problem, at least to most of us.  
Gerry was well known for his physics intuition as well as his brilliant ideas in making complicated
problems simple. His Brown-Rho scaling is a typical example. A main result of the well-known Brown-Rho (BR) 
scaling is  
\cite{brownrho91,hatsuda,brownrho04}  
\begin{eqnarray}
&& \frac{m^*_{\sigma}}{m_{\sigma}}\simeq
\frac{m^*_{N}}{m_{N}}\simeq
\frac{m^*_{\rho}}{m_{\rho}}\simeq
\frac{m^*_{\omega}}{m_{\omega}}\simeq \Phi_{BR} (n), \nonumber \\
&& \Phi_{BR} (n)= 1-C\frac{n}{n_0},
\end{eqnarray}
where $m^*$ and $m$ denote respectively the in-medium and in-vacuum
mass. Here the parameter $C$ has the value $0.15-0.20$, $n$ is the density
of the nuclear medium, and $n_0$ is nuclear matter saturation density ($0.16 fm^{-3}$). It is remarkable 
that the above simple scaling law, derived in the context of chiral symmetry restoration in dense matter, 
would have dramatic consequences for traditional nuclear structure physics. We shall denote the above linear
scaling as the BR scaling. This scaling naturally
renders $V_{NN}$ a density dependent interaction $V_{NN}(n)$. 

Before discussing the various effects of BR scaling, let me (TTSK)
first recall a conversation with Gerry many 
years ago, probably in 1964 when I attended Gerry's Nuclear Physics course at Princeton. 
He talked a lot 
about Brueckner theory and also about the empirical Skyrme effective interaction
\cite{ringschuck80} of the form
\begin{eqnarray}
V_{skyrme}&=&V_{sk}(\vec r_1-\vec r_2)+D_{sk}(\vec r_1-\vec r_2), \nonumber \\ 
D_{sk}&=&\frac{1}{6}(1+x_3P_{\sigma})t_3\delta (\vec r_1 - \vec r_2)
n(\vec r_{av}),
\end{eqnarray}
where $V_{sk}$ is a two-body $\delta$-function force and $D_{sk}$ is a density-dependent two-body interaction.
It was a bit `strange' that there was a piece of interaction 
which was density dependent. 
One day Gerry mentioned that it would be nice to work out a connection
between the empirical Skyrme interaction and meson-exchange NN
interactions. At that time, my understanding of them 
was minimal I have to confess,
and I was really unable to pursue the matter further. 
But now there are, I think,
indications \cite{dong09,dong11} that BR-scaling may provide 
a microscopic foundation
for the density-dependent Skyrme effective interaction. 

We have carried out several studies on the effects
of BR scaling on finite nuclei, nuclear matter and neutron stars
\cite{holtbrs07,holtc1408,siu09,dong09,dong11,dongnewbr13}.
Let us just briefly describe a few of them. For convenience in
implementing BR scaling, we have employed the BonnA and/or BonnS
one-boson-exchange potentials \cite{bonnabc,bonns} whose
parameters for $\rho$, $\omega$ and $\sigma$ mesons are scaled with the density
(in our calculations the meson masses and cut-off parameters are equally scaled). 
Note that $\pi$ is protected by chiral symmetry and is not scaled.
That we scale $\rho$ but not $\pi$ has an important consequence for the
tensor force, which plays an important role in the famous Gamow-Teller (GT) matrix element for the 
$^{14}C\rightarrow$ $^{14}N$ $\beta$-decay
\cite{holtc1408}. The tensor force from $\pi$- and $\rho$-meson exchange are of opposite
signs. A lowering of only $m_{\rho}$, but not $m_{\pi}$, can significantly
suppress the net tensor force strength and thus largely diminish the GT matrix element.
In addition, the scaling of the $\omega$ meson introduces additional short-distance
repulsion into the nucleon-nucleon interaction, which was found to also contribute to
the suppression \cite{holt09}. With BR-scaling we were able to satisfactorily reproduce the 
$\sim 5800$-yr long lifetime of this decay; I remember well that Gerry was
very pleased with this result.

BR scaling has been applied to several nuclear matter calculations 
\cite{holtbrs07,siu09,dong09,dong11,dongnewbr13}.
Let me start from a result, as shown in Fig.\ 4 of Ref.\ \cite{dongnewbr13},
to illustrate the current situation.
This calculation employed a so-called new-Brown-Rho (new-BR) scaling
\cite{dongnewbr13}
which is based on a half-Skyrmion model to be described briefly
later. The equation of state (EOS) labeled (C) is obtained from $V_{low-k}$ (derived
from the BonnS $V_{NN}$ \cite{bonns}) without any BR scaling. It does not exhibit satisfactory saturation
properties. 
It is a general result that $V_{NN}$ alone cannot give satisfactory nuclear saturation
properties as illustrated by (C) of Fig.\ 4.

We have found that Brown-Rho scaling improves this situation dramatically
\cite{holtbrs07,siu09,dong09,dong11,dongnewbr13}.
Moreover the combined potential given by the sum of the
unscaled-$V_{NN}$ and $D_{sk}$, the latter being the Skyrme density-dependent
force of Eq.\ (7), can also give equally satisfactory nuclear matter
saturation properties \cite{dong09}. Four such EOS's are shown in
Fig.\ 5, where $\Lambda$ denotes the decimation momentum scale for $V_{low-k}$.
As seen, the four EOS's agree with each other closely, all giving
$E_0/A \simeq -15$ MeV, $k_F \simeq 1.40$ fm$^{-1}$ and $K \simeq 150$ MeV.
Thus to have satisfactory nuclear matter saturation properties, we may use either a BR-scaled $V_{NN}$ or
an unscaled-$V_{NN}$ + $D_{sk}$; this indicates that a microscopic foundation for the empirical Skyrme 
density-dependent force may be provided by BR scaling.

\begin{figure}[here]
%\scalebox{0.34}{\includegraphics[angle=-90]{11113newbr3.eps}}
%\scalebox{0.36}{\includegraphics[angle=-90]{11113newbr.eps4esym2015}}
\scalebox{0.36}{\includegraphics[angle=-90]{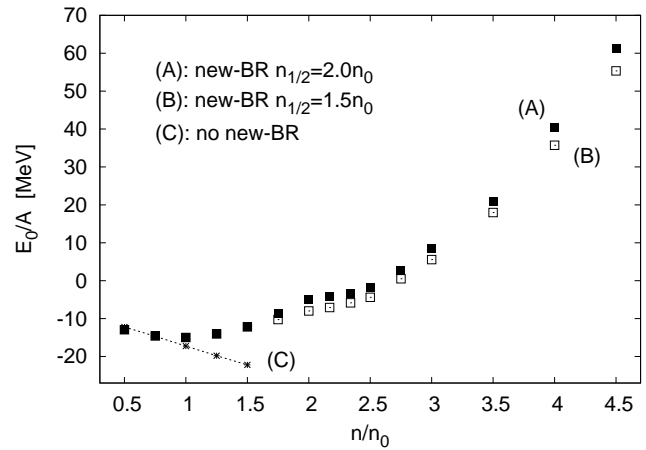}}
\caption{Comparison of the EOS for symmetric nuclear matter calculated with
and without the new-BR scaling. Transition densities of
$n_{1/2}=2.0$  (solid square) and $1.5n_0$ (open square) are employed.
 See text for more explanations.}
\end{figure}

\begin{figure}[here]     
%\scalebox{0.42}{\includegraphics[angle=-90]{818withskyrmesym}}
\scalebox{0.42}{\includegraphics[angle=-90]{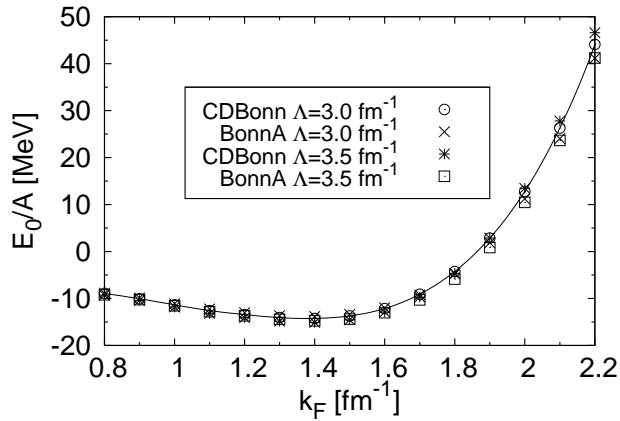}}
\caption{Ring-diagram EOS for symmetric nuclear matter with the interaction
being the sum of  $V_{low-k}$  and the Skyrme density dependent
 force of Eq.(7). 
Four sets of results are shown for CDBonn and BonnA potentials 
with $\Lambda$=3 and
 3.5 $fm^{-1}$. A common Skyrme force of $t_3$=2000 MeV-$fm^6$ 
and $x_3=0$ is 
employed.}
\end{figure}          

We now describe the new-BR scaling \cite{dongnewbr13} 
on which the results shown in Fig.\ 4 are based.
The idea behind this scaling is that when a large number of
skyrmions as baryons are put on an FCC (face-centered-cubic) crystal
to simulate dense
matter, the skyrmion matter undergoes a transition to a matter consisting
of half-skyrmions~\cite{goldhaber} in CC configuration at a density
that we shall denote as $n_{1/2}$. This density is difficult to pin
down precisely but it is more or less independent of the mass of the
dilaton scalar, the only low-energy degree of freedom that is not
well-known in free space. The density at which this occurs has been estimated to lie typically
between 1.3 and 2 times normal nuclear matter density $n_0$
~\cite{half}.
In our model, nuclear matter is separated into two regions I and II
respectively for densities $n\leq n_{1/2}$ and $n>n_{1/2}$. As inferred
by our model, they have different scaling functions
\begin{equation}
\Phi_i (n) = \frac{1}{1+C_i \frac{n}{n_0}},~~ i=I,II.
\end{equation}
The above two-region scaling is the new-BR scaling mentioned earlier. 

The EOS (A) and (B) of Fig.\ (4) are obtained with the new-BR scaling with
 $n_{1/2}$=  1.5 and 2.0$n_0$ respectively. As described in
\cite{dongnewbr13}, we employ in our new-BR calculations
 the BonnS potential \cite{bonns} with scaling parameters
$C_{\rho}$=0.13,
$C_{\sigma}$=0.121,
$C_{\omega}$=0.139 ,
$C_{N}$=0.13 and
$C_{g,\rho}=C_{g,\omega}$=0 for region I.
For region II the scaling parameters are
$C_{\rho}$=0.13,
$C_{\sigma}$=0.121,
$C_{\omega}$=0.139 ,
$C_{g,\rho}$=0.13, $C_{g,\omega}$=0 and $m^*_N/m_N=y(n)=0.77$. 
Note that this scaling has some special features:
In region I the coupling constants $g_{\rho N}$ and $g_{\omega N}$
are not scaled, while in region II only the coupling constant $g_{\rho N}$
is scaled. Also in region II 
the nucleon mass is a density-independent 
constant ($m^* _N / m_N$=0.77). 
Note that our choices
for the $C$ parameters are consistent with the Ericson scaling
which is based on a scaling relation for the quark condensate
$\frac{<\bar qq>^*}{<\bar qq>}$ \cite{ericson}. According to this
scaling, at low densities one should have $C\simeq D/3$ with
$D=0.35 \pm 0.06$.

 We note that the above scaling is only `inferred'  by our
Skyrmion-half-Skyrmion model \cite{dongnewbr13}.
As a first step to check this scaling, we have carried 
out several applications.
In Fig.\ 4, the calculated EOS for symmetric nuclear matter
using transition densities $n_{1/2}$= 1.5 (A) and 2.0$n_0$ (B)
are shown. Both give an energy per nucleon 
$E_0/A=-15$ MeV, saturation density $k_F=1.30 fm^{-1}$
 and compression modulus $K$=208 MeV, all in satisfactory
agreement with the empirical values \cite{dongnewbr13}.
We believe that our scaling works 
well for low densities of $n \lesssim 1.5n_0$.

 How to scale the mesons at densities beyond $n_0$ is still
an open question.
 By way of heavy-ion collision experiments,
there has been much progress  in determining
the nuclear symmetry energy $E_{sym}$ up to densities as high
as $\sim 5n_0$
\cite{li05,li08,tsang09}.
Thus an application of our new-BR  scaling
to the calculation of $E_{sym}$ would provide an important test
for this scaling in the region with $n>n_{1/2}$.
As displayed in Fig.\ 6,  our calculated symmetry
energies agree well  with the experimental constraints 
\cite{li05,tsang09}.

\begin{figure}[here]
%\scalebox{0.34}{\includegraphics[angle=-90]{12113newbr9.eps}}
%\scalebox{0.36}{\includegraphics[angle=-90]{90112newbr.epsesym}}
\scalebox{0.34}{\includegraphics[angle=-90]{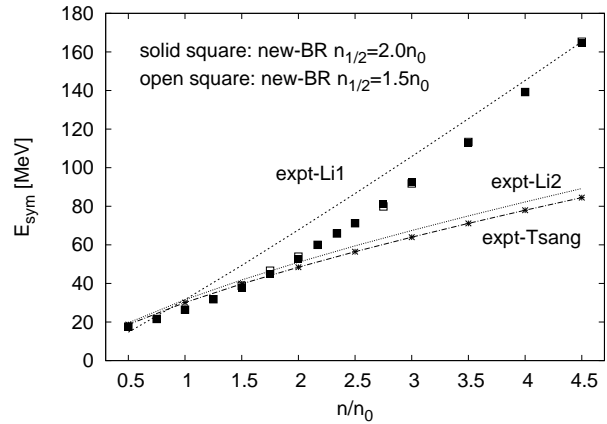}}
\caption{
Comparison of our calculated
nuclear symmetry energies
with the empirical upper (expt-Li1)
and lower (expt-Li2) constraints
of Li et al. \cite{li05} and the empirical results of
Tsang et al. (expt-Tsang) \cite{tsang09}.
 }
\end{figure}

The EOS at high densities ($n \simeq 5-10n_0$) is important for
neutron-star properties. Thus an application of the new-BR scaling
to neutron star structure would provide a useful test. As shown in Fig.\ 7,
our calculated neutron-star maximum mass is about $2.4 M_{\odot}$,
 slightly larger than the empirical value of $\sim 2M_{\odot}$
\cite{dongnewbr13}.
In our calculations, the central
densities of neutron stars are $\sim 5n_0$. At such densities,
 how to scale the hadrons
with the medium remains to be an interesting and open question.
Much remains to be done. 

\begin{figure}[here]
%\scalebox{0.34}{\includegraphics[angle=-90]{12113newbr9.eps}}
\scalebox{0.33}{\includegraphics[angle=-90]{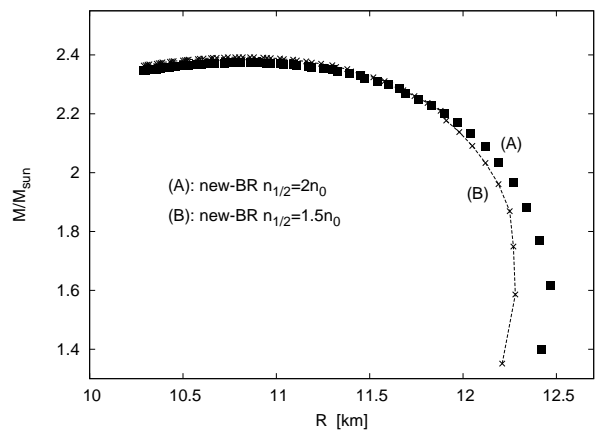}}
\caption{Mass-radius trajectories of
neutron stars calculated
with new-BR scalings using $n_{1/2}$ = 2.0 (A) and 1.5$n_0$ (B).
The maximum neutron-star
mass and its radius for these two cases are
respectively \{$M_{\rm max} = 2.39 M_{\odot}$ and $R = 10.90$ km\} and
\{ $M_{\rm max} = 2.38 M_{\odot}$ and $R= 10.89$ km\}. }
\end{figure}

\section{IV. Summary}
It is indeed our fortune and privilege to have met Gerry and worked with 
him since our early careers. He was not only a great scientist but also a 
kind person, who supported his colleagues and students over many years. I 
(TTSK) especially have many fond memories of fun times together, starting 
in 1964 when I met Gerry at Princeton, and our relationship extended beyond 
academics (such as our regular tennis matches). Gerry's insights into core 
polarization and Brown-Rho scaling are of fundamental
importance for our understanding of effective nucleon-nucleon 
interactions in a
nuclear medium and the possible connections to chiral 
symmetry restoration in dense matter.
It was a pleasure to explore these topics together, and we have 
learned much from 
him over the years. Let us express our deep gratitude to him.

\vskip .2in

{\bf Acknowledgement} We are grateful to M.\ Rho and R.\ Machleidt 
for helpful discussions. This work was supported in part
by the Department of Energy under Grant No.\ DE-FG02-88ER40388
and DE-FG02-97ER-41014.

\end{document}